\documentclass[12pt,draftclsnofoot,onecolumn]{IEEEtran}
\usepackage {amssymb,rotating,graphicx,cite, calc, color, psfrag}
\usepackage{authblk}
\usepackage[cmex10]{amsmath}
\usepackage{amssymb,amsthm,amsfonts,mathrsfs}
\usepackage{graphicx}
\usepackage{enumitem,stackrel}
\usepackage{color}
\usepackage{mathtools,amssymb,bm}
\usepackage{amstext}
\usepackage{array}
\usepackage{algorithm}
\usepackage{algorithmicx}
\usepackage{epsfig}

\usepackage{hyperref}
\theoremstyle{remark}

\ifCLASSOPTIONcaptionsoff
  \usepackage[nomarkers]{endfloat}
 \let\MYoriglatexcaption\caption
 \renewcommand{\caption}[2][\relax]{\MYoriglatexcaption[#2]{#2}}
\fi
\newcommand{\RN}[1]{%
	\textup{\uppercase\expandafter{\romannumeral#1}}%
}

\def\bphi{\boldsymbol{\phi}}

\newtheoremstyle{mystyle}
  {}
  {}
  {\itshape}
  {}
  {\bfseries}
  {.}
  { }
  {}
\theoremstyle{mystyle}

{}
{}
{}

{}
{}
{}
{}
{}
\usepackage{epstopdf}
\usepackage{cuted}
\usepackage{flushend}
\usepackage{enumitem}
\usepackage{amsfonts}
\usepackage{gensymb}
\usepackage[english]{babel}
\usepackage[utf8]{inputenc}
\usepackage[noend]{algpseudocode}
\usepackage{amsmath}
\setlist[itemize,1]{label=$\times$}
\setlist[itemize,2]{label=$\checkmark$}
\setlist[itemize,3]{label=$\diamond$}
\setlist[itemize,4]{label=$\bullet$}
\usepackage{graphicx}                  
\usepackage{fancyhdr}                  

\begin{document}
\title{Vertical Beamforming in Reconfigurable Intelligent Surface-aided Cognitive Radio Networks}
\author{S. Fatemeh~Zamanian, S. Mohammad~Razavizadeh, and Qingqing~Wu
\thanks{S. Fatemeh Zamanian and S. Mohammad~Razavizadeh are with the School of Electrical Engineering, Iran University of Science \& Technology (IUST), Tehran
1684613114, Iran (e-mail: f\underline\;zamanian@elec.iust.ac.ir; smrazavi@iust.ac.ir).

Qingqing~Wu is with the State Key Laboratory of Internet of Things for Smart City, University of Macau, China (e-mail: qingqingwu@um.edu.mo).
}}%
\maketitle
\begin{abstract}
In this letter, we investigate joint application of reconfigurable intelligent surface (RIS) and vertical beamforming in cognitive radio networks (CRN). After properly modeling the network, an optimization problem is formed to jointly design the beamforming vector and tilt angle at the secondary base station (BS) as well as the phase shifts at the RIS with the objective of maximizing spectral efficiency of the secondary network. The optimization problem is non-convex; thus, we propose an efficient solution method for it. Numerical results show that adding a RIS and optimizing the radiation orientation, can significantly improve performance of the CRNs.
\end{abstract}
\IEEEpeerreviewmaketitle
\section{Introduction}\label{Introduction}
Recently, reconfigurable intelligent surface (RIS), also known as intelligent reflecting surface (IRS), has been considered as a key technology to manipulate the wireless propagation environment for achieving various objectives \cite{1}. An RIS-aided network comprises a programmable meta-surface with massive reflecting elements that their phases are optimized in a way to improve some metrics such as interference reduction, security enhancement, and energy efficiency \cite{3} - \cite{05}.  

On the other hand, another well-known technique  for improving the spectrum usage in wireless channels is cognitive radio which has always been  known as a promising candidate for evolving the wireless networks \cite{13_new}. RIS technology can be used in the cognitive radio networks (CRN) for further improvement in spectrum efficiency. In \cite{06}, the authors maximized the achievable weighted sum rate of the  secondary system in an RIS-aided multiple-input multiple-output (MIMO) CRN wherein the precoding vector of the secondary base station (SBS) and the phase shifts of the RIS were jointly optimized.
Authors in \cite{60} and \cite{6} maximized the achievable rate of the secondary system in a single-RIS-assisted downlink multiple-input single-output (MISO) CRN and in a multi-RIS-assisted downlink MISO CRN through joint optimization of the beamforming of the secondary transmitter and the RIS’s phases. A similar optimization approach was proposed in \cite{70} - \cite{2} to address the resource allocation and spectrum sharing problems in the RIS-assisted CRNs. Moreover, in \cite{8}, it was illustrated how to equip CRNs with RIS to solve the security issue attributed to CRNs.

In addition to the above technologies, three dimensional (3D) beamforming is another evolving technology in new generations of wireless networks in which the radiation pattern of the base station (BS) in the elevation and azimuth domains are carefully adjusted to improve signal reception at some desired locations \cite{12_new}. Because of the low sensitivity of the radiation pattern to the azimuth angle, 3D beamforming usually leads to optimization of the tilt angle, and, therefore, is also known as the vertical beamforming \cite{111}. Particularly, vertical beamforming is useful for improving different network metrics such as spectral and energy efficiencies and security \cite{5}. Vertical beamforming can also be jointly used with the RIS technology for more improvement in the network performance. The authors in \cite{9} contemplated a BS with 3D beamforming capability to provide more degrees of freedom in design and deployment of the RIS-assisted MISO networks.

In this letter, we show how the performance of a CRN is improved by equipping it with an RIS and vertical beamforming mechanism at the SBS. In this way, we first propose a system model including a primary base station (PBS), an SBS and an RIS that helps both BSs in their signal transmissions. While both BSs are equipped with multiple antennas, only the SBS has the capability of vertical beamforming. Then, we aim to maximize the spectral efficiency (SE) of the secondary network by jointly optimizing the beamforming vector and tilt angle at the SBS and the phase shifts at the RIS. 
The proposed joint optimization problem is non-convex and hence we propose an efficient method to solve it. Specifically, we utilize alternative optimization and semidefinite relaxation (SDR) techniques to iteratively optimize the variables. We further extend the results to the 3D beamforming case. Also, we calculate the complexity of our proposed method. Numerical results illustrate that the performance of the CRNs can considerably improve when the tilt angle radiates in the direction of the RIS, and also the SE of the system gets better when the number of reflecting elements of the RIS increases.
\begin{figure}[!h]
\centering
\begin{minipage}[b]{0.7\textwidth}
   \includegraphics[width=\textwidth]{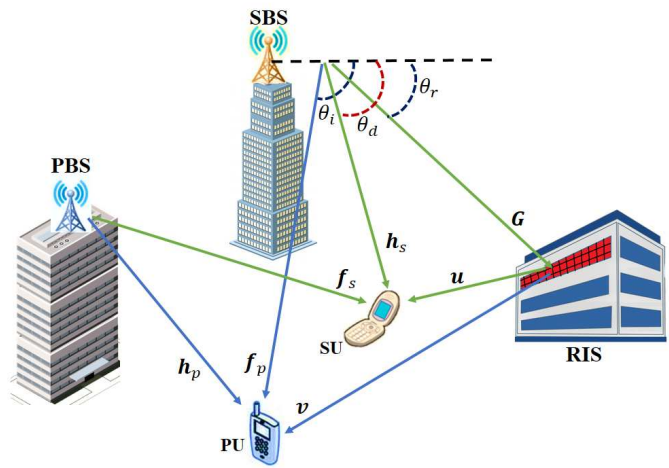}
 \caption{\label{system2} System Model of the RIS-aided cognitive radio network.}
\end{minipage}
\end{figure}
\section{SYSTEM MODEL}\label{SYSTEM_MODEL}
As illustrated in Fig. \ref{system2}, we consider the downlink transmission of an MISO CRN using underlay spectrum sharing. The network consists of a primary and a secondary system. The primary system consists of a PBS equipped with $N_p$ antennas that serves a single-antenna primary user (PU). Also, the secondary system consists of an SBS equipped with $N_s$ antennas that serves a single-antenna secondary user (SU). Furthermore, an RIS comprising $N$ reflecting elements is deployed to assist both primary and secondary transmissions. The SBS is equipped with a full-dimensional array of antennas that adopts an optimized beamforming vector and tilt angle to transmit signals to its intended receivers by the help of the RIS. Besides, we only consider the first-order reflection from the RIS due to the significantly path loss.

In Fig. \ref{system2}, ${\bf{G}}\in \mathbb{C}^{N\times N_s}$ denotes the channel matrix between the  SBS and the RIS. Also, ${\bf{v}}\in \mathbb{C}^{N\times 1}$ and ${\bf{u}}\in \mathbb{C}^{N\times 1}$ are the channel vectors between the RIS and the PU and SU, respectively.
In addition, ${\bf{h}}_p\in \mathbb{C}^{N_p\times 1}$ and ${\bf{h}}_s\in \mathbb{C}^{N_s\times 1}$ denote the channel vectors between the PBS and SBS and the PU and SU, respectively.  Moreover, ${\bf{f}}_p\in \mathbb{C}^{N_s\times 1}$ and ${\bf{f}}_s\in \mathbb{C}^{N_p\times 1}$ are the interference channel vectors between the SBS and PBS and the PU and SU, respectively.

Note that to employ vertical beamforming, we model the vertical antenna attenuation (pattern) at the SBS as follows \cite{9}
\begin{align}
a_V^x( \theta_{tilt}, \theta_x) =-min\left[12 \left(\frac{\theta_x-\theta_{tilt}}{\theta_{3dB}}\right)^2 , SLA_V\right],
\end{align}
where $x\in\{d, r, i\}$ and $\theta_d$, $\theta_r$ and $\theta_i$ are the elevation angles of the SU, RIS and PU, respectively. Moreover, $\theta_{tilt}$ is the vertical tilt angle, $\theta_{3dB}$ is the vertical 3 dB beamwidth, and $SLA_V$ is the maximum side-lobe level where it is usually assumed that $SLA_V=\infty$. Since $\theta_x$ is fixed for given positions for the users and RIS, we define the vertical antenna attenuation (pattern) as $a_V^x( \theta_{tilt})$, or in linear scale as $A_V^x( \theta_{tilt})$. Therefore, the received signal at the RIS can be written as
\begin{align} \label{002}
{\bf{r}}_r=\sqrt{A_V^r(\theta_{tilt})}{\bf{\Phi}}{\bf{G}}{\bf{w}}_ss_s,
\end{align}  
where ${\bf{\Phi}}\triangleq \rm{diag} ({{\bphi}})=\rm{diag} ({\phi_1},{\phi_2},...,{\phi_N})$ is a diagonal matrix accounting for the effective phase shifts applied by all passive RIS reflecting elements, where $\phi_n=e^{j\alpha_n},\;\forall n=1,...,N$\footnote {To maximize the signal reflection of the RIS, the reflection amplitude is set as one \cite{1}.}. Moreover, ${\bf{w}}_s$ is beamforming vector at the SBS. Also, $s_s$ denotes the normalized data signal transmitted by the SBS.

After reflecting from the RIS, the received signals at the SU and PU are as follows, respectively
\begin{align} \label{2}
r_s=\underbrace{\sqrt{A_V^d(\theta_ {tilt})}{\bf{h}}_s^H{\bf{w}}_ss_s}_{\rm{direct\;link}}+\underbrace{{\bf{f}}_s^H{\bf{w}}_ps_p}_{\rm{\scriptstyle interference\;\hfill\atop\scriptstyle \;\;\;\;\; link\hfill}}+\underbrace{{\bf{u}}^H{\bf{r}}_r}_{\rm{\scriptstyle reflected\;from\hfill\atop\scriptstyle \;\;\;\;\;\;RIS\hfill}}+n_s,
\end{align}
\begin{align} \label{1}
r_p=\underbrace{{\bf{h}}_p^H{\bf{w}}_ps_p}_{\rm{direct\;link}}+\underbrace{\sqrt{A_V^i(\theta_{tilt}) }{\bf{f}}_p^H{\bf{w}}_ss_s}_{\rm{\scriptstyle interference\;\hfill\atop\scriptstyle \;\;\;\;\; link\hfill}}+\underbrace{ {\bf{v}}^H{\bf{r}}_r}_{\rm{\scriptstyle reflected\;from\hfill\atop\scriptstyle \;\;\;\;\;\;RIS\hfill}}+n_p,
\end{align}
where  ${\bf{w}}_p$ is beamforming vector at the PBS. Furthermore, $s_p$ denotes the normalized data signal transmitted by the PBS.
 Also, $n_s$, $n_p \in {\cal C}{\cal N}\left( {0,\sigma_n^2} \right)$ signify circularly-symmetric complex Gaussian (CSCG) noise at the SU and PU, respectively.

In this paper, we aim to maximize the SE of the secondary system. To this end, we first derive the SE of the secondary system in terms of the network parameters including the SBS tilt angle, beamforming vector and RIS phase shifts, and then maximize the SE through an optimization problem. 

 Specifically, based on (\ref{2}), the signal-to-interference-plus-noise ratio (SINR) of the SU can be expressived as 
\begin{align} \label{3}
SINR_s=\frac{\left|\left(\sqrt{A_V^d(\theta_ {tilt})}{\bf{h}}_s^H+\sqrt{A_V^r(\theta_{tilt})}{\bf{u}}^H{\bf{\Phi}}{\bf{G}}\right){\bf{w_s}}\right|^2}{\sigma_n^2+|{\bf{f}}_s^H{\bf{w}}_p|^2}.
\end{align}
Then, the secondary SE  is obtained as
\begin{align} \label{4}
SE_s=\log_2\left( 1+ SINR_s\right).
\end{align}
According to the above, to maximize the $SE_s$, we formulate an optimization problem as follows
\begin{align} \label{5}
\begin{array}{l}
\!\!\!\mathop {\max }\limits_{{\bf{\Phi}},\;\theta_{tilt},\;\bf{w}_s} \log_2\left( 1+\frac{\left|\left(\sqrt{A_V^d(\theta_ {tilt})}{\bf{h}}_s^H+\sqrt{A_V^r(\theta_{tilt})}{\bf{u}}^H{\bf{\Phi}}{\bf{G}}\right){\bf{w}}_s\right|^2}{\sigma_n^2+|{\bf{f}}_s^H{\bf{w}}_p|^2} \right)\\
{s.t.}\;
{C_1}:\left|\left(\sqrt{A_V^i(\theta_{tilt}) }{\bf{f}}_p^H+ \sqrt{A_V^r(\theta_{tilt}) }{\bf{v}}^H{\bf{\Phi}}{\bf{G}}\right){\bf{w}}_s\right|^2 \le \Gamma,\\
\;\;\;\;\;\;{C_2}: |\phi_n|^2=1,\;\;\;\forall n=1,...,N,\\
\;\;\;\;\;\;{C_3}:-{\pi} \le \theta_{tilt} \le 0,\\
\;\;\;\;\;\;{C_4}:\sum\limits_{i=1}^{N_s} |w_{s_i}|^2\le P.
\end{array}
\end{align}
The constraints of (\ref{5}) are as follows.
 $C_1$ guarantees the primary system interference condition, where ${\rm{\Gamma }}$ is the interference threshold of the primary network.
 Furthermore, $C_2$ is the RIS-gain constraint.
Moreover, $C_3$ denotes the allowable interval of each  $\theta_{tilt}$.
Also, $C_4$ represents the power budget condition of the SBS where $P$ is the maximum total power of the SBS. 

The optimization variables in the objective function of problem (\ref{5}) are coupled so that the objective function exhibits a non-convex form. Moreover, the left-hand-side of the constraint C1 is a non-convex function and also the left hand side of the equality constraint C2 is nonlinear. Therefore, the problem (\ref{5}) is a highly non-convex problem and it is difficult to find its optimal value in polynomial time. Accordingly, to solve it, we utilize SDR technique and propose an alternating method as follows.
\section{Proposed Method}\label{pf}
In this section, we propose  an efficient solution method for (\ref{5}).
By ignoring the monotonic logarithm function and constant terms, we have an equivalent optimization problem as follows
\!\!\begin{align} \label{6}
\begin{array}{l}
\mathop {\max }\limits_{{\bf{\Phi}},\;\theta_{tilt},\;\bf{w}_s}\; {\left|\left(\sqrt{A_V^d(\theta_ {tilt})}{\bf{h}}_s^H+\sqrt{A_V^r(\theta_{tilt})}{\bf{u}}^H{\bf{\Phi}}{\bf{G}}\right){\bf{w}}_s\right|^2}\\
{s.t.}\;
{C_1}:\left|\left(\sqrt{A_V^i(\theta_{tilt}) }{\bf{f}}_p^H+ \sqrt{A_V^r(\theta_{tilt}) }{\bf{v}}^H{\bf{\Phi}}{\bf{G}}\right){\bf{w}}_s\right|^2 \le \Gamma,\\
\;\;\;\;\;\;{C_2}: |\phi_n|^2=1\;\;\;\forall n=1,...,N,\\
\;\;\;\;\;\;{C_3}:-{\pi} \le \theta_{tilt} \le 0,\\
\;\;\;\;\;\;{C_4}:\sum\limits_{i=1}^{N_s} |w_{s_i}|^2\le P.
\end{array}
\end{align}
In the following, we first obtain the optimum value of  $\theta_{tilt}$ with given ${\bf{\Phi}}$ and ${\bf{w}}_s$ and then optimize ${\bf{w}}_s$ with given ${\bf{\Phi}}$ and $\theta_{tilt}$. After that, we obtain the optimum value of  ${\bf{\Phi}}$ with given $\theta_{tilt}$ and ${\bf{w}}_s$, and we propose an alternating algorithm to solve (\ref{6}). Finally, we analyze the computational complexity of our proposed method.
\subsection{Optimizing  $\theta_{tilt}$ with Given ${\bf{\Phi}}$ and ${\bf{w}}_s$}\label{753}
 We firstly focus on the objective function of (\ref{6}) to obtain the optimum value of  $\theta_{tilt}$, i.e. $\theta_{tilt}^*$. Let us rewrite the objective function of (\ref{6}) as follows
\begin{align}\label{15}
{\left|10^{-0.6(\frac{\theta_d-\theta_{tilt}}{\theta_{3dB}})^2}{\bf{h}}_s^H{\bf{w}}_s+10^{-0.6(\frac{\theta_r-\theta_{tilt}}{\theta_{3dB}})^2}{\bf{u}}^H{\bf{\Phi}}{\bf{G}}{\bf{w}}_s\right|^2}.
\end{align}
\;\;\;It is not difficult to check that the curve of $10^{-0.6(\frac{\theta_x-\theta_{tilt}}{\theta_{3dB}})^2}$ has a unique maximum value at $\theta_x$. Therefore, by assuming $\theta_r\neq\theta_d$, (\ref{15}) has two extrema at $\theta_d$ and $\theta_r$. Now, if $|{\bf{h}}_s^H{\bf{w}}_s| > |{\bf{u}}^H{\bf{\Phi}}{\bf{G}}{\bf{w}}_s|$, then $\theta_{tilt}^*=\theta_d$, and if  $|{\bf{h}}_s^H{\bf{w}}_s| <|{\bf{u}}^H{\bf{\Phi}}{\bf{G}}{\bf{w}}_s|$, then $\theta_{tilt}^*=\theta_r$.\\
Due to the randomness of the channels, we compare the expectation of $|{\bf{h}}_s^H{\bf{w}}_s|^2$ and $|{\bf{u}}^H{\bf{\Phi}}{\bf{G}}{\bf{w}}_s|^2$ to find $\theta_{tilt}^*$. By assuming that all channels' elements have identically independent distributions of ${\cal C}{\cal N}\left( {0,\sigma^2}\right)$, we have
\begin{align}\label{16}
E\left[|{\bf{h}}_s^H{\bf{w}}_s|^2|\right]\mathop {= }\limits^{{a}}\sigma^2 Tr({\bf{w_s}}{\bf{w_s}}^H)=\sigma^2||{\bf{w_s}}||^2_2, 
\end{align}
where $Tr(\cdot)$ stands for trace of matrix and $a$ is because of the following equalities
\begin{align}\label{TraceExpect}
\begin{array}{l}
E\left[Tr\left(.\right)\right]=Tr\left(E\left[.\right]\right), \\
Tr\left({\bf{A}}{\bf{B}}\right)=Tr\left({\bf{B}}{\bf{A}}\right).
\end{array}
\end{align}
Also
\begin{align}\label{17}
&E\left[|{\bf{u}}^H{\bf{\Phi}}{\bf{G}}{\bf{w}}_s|^2\right]\mathop {= }\limits^{{b}}\sigma^2 Tr\left(E\left[{\bf{\Phi}}{\bf{G}}{\bf{w}}_s{\bf{w}}_s^H{\bf{G}}^H {\bf{\Phi}}^H\right]\right)\nonumber\\&\mathop {= }\limits^{{c}}\sigma^2 E\left[Tr\left({\bf{G}}^H{\bf{G}}{\bf{w}}_s{\bf{w}}_s^H\right)\right]\mathop {= }\limits^{{d}}\sigma^2 Tr\left(E\left[{\bf{G}}^H{\bf{G}}\right]{\bf{w}}_s{\bf{w}}_s^H\right)\nonumber\\ &=\sigma^4N||{\bf{w_s}}||^2_2,
\end{align}
where $b$ is verified by (\ref{TraceExpect}) and also independent of ${\bf{u}}$ and ${\bf{G}}$, $c$ is due to ${\bf{\Phi}}^H{\bf{\Phi}}={\bf{I}}_N$ and $d$ is due to ${\bf{w}}_s$ is deterministic in this subsection.

When $\sigma^4N||{\bf{w_s}}||^2_2> \sigma^2||{\bf{w_s}}||^2_2$, i.e. $\sigma^2N>1$, we obtain $\theta_{tilt}^*=\theta_r$. This is the case for large values of $N$, which is usual in RIS technology. Moreover, for sufficiently large value of $\Gamma$, constraint $C1$ of (\ref{6}) will be satisfied for $\theta_{tilt}^*=\theta_r$. Our simulation results reveal that the value of $\Gamma=1$ w is enough.

It should be noted that, extending our results from tilt angle optimization to 3D beamforming, i.e. tilt and azimuth angles optimization, is straightforward. To elaborate, consider the overall SBS antenna gain in linear scale as 
\begin{align}\label{18}
A^x(\theta_{tilt},\phi_{azimuth})=A_m10^{-1.2\left[\left(\frac{\theta_x-\theta_{tilt}}{\theta_{3dB}}\right)^2+\left(\frac{\phi_x-\phi_{azimuth}}{\phi_{3dB}}\right)^2\right]},
\end{align}
where $x\in\{d, r, i\}$ and $\phi_d$, $\phi_r$ and $\phi_i$ are the horizontal angles of the SU, RIS and PU, respectively. Moreover, $\phi_{azimuth}$ is the azimuth angle, $\phi_{3dB}$ is the horizontal 3 dB beamwidth, and $A_m$ is the maximum directional gain of the antenna array elements. 
It is easy to verify optimizing $\phi_x$ is completely similar to that of $\theta_x$ and the optimal $\phi_x$ is $\phi_r$. Therefore, the spectral efficiency of the secondary system improves when the tilt and azimuth angles orient towards the RIS.
\subsection{Optimizing ${\bf{w}}_s$ with Given ${\bf{\Phi}}$ and $\theta_{tilt}$}\label{741}
By fixing ${\bf{\Phi}}$ and $\theta_{tilt}$, we need to solve the following sub-problem. 
\begin{align} \label{7}
\begin{array}{l}
\mathop {\max }\limits_{\bf{w_s}}\;\;\;\; {\bf{a}}{\bf{w}}_s{\bf{w}}_s^H{\bf{a}}^H\\
{s.t.}\;\;\;\;
 {\bf{b}}{\bf{w}}_s{\bf{w}}_s^H{\bf{b}}^H\le \Gamma,\\
\;\;\;\;\;\;\;\;\;\sum\limits_{i=1}^{N_s} |w_{s_i}|^2\le P,
\end{array}
\end{align}
where ${\bf{a}}=\sqrt{A_V^d(\theta_ {tilt})}{\bf{h}}_s^H+\sqrt{A_V^r(\theta_{tilt})}{\bf{u}}^H{\bf{\Phi}}{\bf{G}}$  \;and\; ${\bf{b}}=\sqrt{A_V^i(\theta_{tilt}) }{\bf{f}}_p^H+ \sqrt{A_V^r(\theta_{tilt}) }{\bf{v}}^H{\bf{\Phi}}{\bf{G}}$. Then, we define ${\bf{W}}_s={\bf{w}}_s{\bf{w}}_s^H$ and therefore, the optimization problem can be rewritten as
\begin{align} \label{8}
\begin{array}{l}
\mathop {\max }\limits_{\bf{W_s}}\;\;\;\; {\bf{a}}{\bf{W}}_s{\bf{a}}^H\\
{s.t.}\;\;\;\;
 {\bf{b}}{\bf{W}}_s{\bf{b}}^H\le \Gamma,\\
\;\;\;\;\;\;\;\;\;{\rm{tr}}({\bf{W}}_s)\le P,\\
\;\;\;\;\;\;\;\;\;{\bf{W}}_s \succeq {\bf{0}},\\
\;\;\;\;\;\;\;\;\;rank({\bf{W}}_s)=1.
\end{array}
\end{align}
This optimization problem is still non-convex. Thus, we use SDR technique by  removing $rank({\bf{W}}_s)=1$ to transform it to a convex problem. 
 Then, to address the relaxed constraint $rank({\bf{W}}_s) = 1$, and to obtain solution to problem (\ref{8}), we apply the sequential rank-one constraint relaxation (SROCR) technique \cite{14}.

\begin{algorithm}
\caption{Proposed Solution Method}
\small{
\begin{algorithmic}[1]
{}{}\\
  {\bf{Requirement:}} $\theta_r, \theta_d, \theta_i, N, N_s, P, \Gamma, \sigma_n$.\\
    {\bf{Initialization:}} ${\bf{\Phi}^{(0)}}$.
\State $\theta_{tilt}^*=\theta_r$.
\State $Err=\infty$.
    \While{$Err \ge \epsilon$}  
        \State  Set $t= t+1$. 
        \State {With given ${\bf{\Phi}}^{(t-1)}$ solve problem (\ref{8}), then apply SROCR technique over its solution to obtain ${\bf{w}}_s^{(t)}$.}
         \State With given ${\bf{w}}_s^{(t)}$ solve problem (\ref{11}), then apply SROCR technique over its solution to obtain $ {\bf{\Phi}}^{(t)}$.    
          \State Obtain $SE_s^{(t)}$ using (\ref{4}).
           \State $Err=\frac{SE_s^{(t)}-SE_s^{(t-1)}}{SE_s^{(t)}}$. 
        \EndWhile  
\end{algorithmic}}
\end{algorithm}
\subsection{Optimizing ${\bf{\Phi}}$ with Given ${\bf{w}_s}$ and $\theta_{tilt}$}\label{7455}
In this section, by using ${\bf{u}}^H{\bf{\Phi}}{\bf{G}}={{\bphi}}^H\;\rm{diag}({\bf{u}}){\bf{G}}$ and ${\bf{v}}^H{\bf{\Phi}}{\bf{G}}={{\bphi}}^H\;\rm{diag}({\bf{v}}){\bf{G}}$ \cite{3}, we have the following sub-problem
\begin{align} \label{10}
\begin{array}{l}
\mathop {\max }\limits_{\bf{x}}\;\;\;\; l_1+{\bf{x}}^H{\bf{H}}_1{\bf{x}}\\
{s.t.}\;\;\;\;
  l_2+{\bf{x}}^H{\bf{H}}_2{\bf{x}}\le \Gamma,\\
\;\;\;\;\;\;\;\;\;\rm{diag}({\bf{x}}{\bf{x}}^H)={\bf{1}},
\end{array}
\end{align}
where ${\bf{x}}=[\bphi^H\;, 1]^H$, $l_1=A_V^d(\theta_ {tilt}){\bf{h}}_s^H{\bf{w}}_s{\bf{w}}_s^H{\bf{h}}_s$ and $l_2=A_V^i(\theta_ {tilt}){\bf{f}}_p^H{\bf{w}}_s{\bf{w}}_s^H{\bf{f}}_p$. Also, ${\bf{H}}_1$ and ${\bf{H}}_2$ are given in (\ref{011}) and (\ref{12}), respectively.
%
\begingroup\makeatletter\def\f@size{10}\check@mathfonts
\begin{equation}\label{011}
{\bf{H}}_1= \begin{bmatrix}
 A_V^r(\theta_ {tilt})\rm{diag}({\bf{u}}){\bf{G}}{\bf{W}}_s{\bf{G}}^H\rm{diag}({\bf{u}})&\sqrt{ A_V^r(\theta_ {tilt}) A_V^d(\theta_ {tilt})}\rm{diag}({\bf{u}}){\bf{G}}{\bf{W}}_s{\bf{h}}_s \\
 \sqrt{ A_V^r(\theta_ {tilt}) A_V^d(\theta_ {tilt})}{\bf{h}}_s^H{\bf{W}}_s{\bf{G}}^H\rm{diag}({\bf{u}})& 0  
\end{bmatrix} 
\end{equation}
\begin{equation}\label{12}
{\bf{H}}_2= \begin{bmatrix}
 A_V^r(\theta_ {tilt})\rm{diag}({\bf{v}}){\bf{G}}{\bf{W}}_s{\bf{G}}^H\rm{diag}({\bf{v}})&\sqrt{ A_V^r(\theta_ {tilt}) A_V^i(\theta_ {tilt})}\rm{diag}({\bf{v}}){\bf{G}}{\bf{W}}_s{\bf{f}}_p \\
 \sqrt{ A_V^r(\theta_ {tilt}) A_V^i(\theta_ {tilt})}{\bf{f}}_p^H{\bf{W}}_s{\bf{G}}^H\rm{diag}({\bf{v}})& 0  
\end{bmatrix} 
\end{equation}
\endgroup

Problem (\ref{10}) is a non-convex problem, thus, we use semidefinite programming to solve it as follows.
\begin{align} \label{11}
\begin{array}{l}
\mathop {\max }\limits_{\bf{X}}\;\;\;\; l_1+\rm{tr}({\bf{H}}_1{\bf{X}})\\
{s.t.}\;\;\;\;
  l_2+\rm{tr}({\bf{H}}_2{\bf{X}})\le \Gamma,\\
\;\;\;\;\;\;\;\;\;\rm{diag}({\bf{X}})={\bf{1}},\\
\;\;\;\;\;\;\;\;\;{\bf{X}} \succeq {\bf{0}},\\
\;\;\;\;\;\;\;\;\;rank({\bf{X}})=1,
\end{array}
\end{align}
where ${\bf{X}}={\bf{x}}{\bf{x}}^H$. Note that, $rank({\bf{X}})=1$ in problem (\ref{11}) is a non-convex constraint. Thus, we use SDR by simply removing $rank({\bf{X}})=1$ to convexify (\ref{11}). After that,  we use SROCR to obtain an approximate solution to problem (\ref{11}).
\subsection{Complexity of the Proposed Solution Method}\label{complexity}
The main complexity of Algorithm 1 is determined by steps 7 and 8. The complexity of these steps are ${\cal{O}}((N_s+1)^{4.5})$ and ${\cal{O}}((N+1)^{4.5})$, respectively. Thus, the complexity of Algorithm 1 is approximately of ${\cal{O}}\left(M\left((N_s+1)^{4.5}+(N+1)^{4.5}\right)\right)$, where $M$ indicates the iteration number required for achieving convergence. Based on our simulations, $M$ is usually less than $4$ for an accuracy of $\epsilon=10^{-3}$.

\begin{table}[!h]
\caption{Parameters} 
\centering 
\small
\begin{tabular}{|l|c|} 
\hline 
Parameter & Value\\ [0.2ex] 
\hline 
Elevation angle of the SU ($\theta_d$) &  -80\degree\\[0.2ex] 
Elevation angle of the RIS ($\theta_r$) & -30\degree\\[0.2ex] 
Elevation angle of the PU ($\theta_i$) & -110\degree  \\[0.2ex] 
Vertical 3 dB beamwidth ($\theta_{3dB}$) & 10\degree  \\[0.2ex] 
Maximum total power of the PBS ($P_p$)& 5 dBw\\[0.2ex] 
Interference threshold of the primary system ( ${\rm{\Gamma }}$) & 1w\\[0.2ex]   
\hline 
\end{tabular}
\label{table:nonlin} 
\end{table}

\section {Numerical Results} \label{NUMERICAL_RESULTS}
In this section, the performance of the proposed scheme is investigated. The parameters that we use are presented in Table I. 
We assume that the SBS, PBS, SU, PU, and RIS are located at $(0,0,30)$, $(100,0,30)$, $(60,20,3)$, $(40,40,3)$, $(0,100,20)$ in meter, in a three dimensional plane, respectively. We have generated all the channel coefficients $\bf{G}$, $\bf{u}$, $\bf{v}$, ${\bf{h}}_p$, ${\bf{h}}_s$, ${\bf{f}}_p$ and ${\bf{f}}_s$ using the relationship $\sqrt {{\zeta _0}{{\left( {{{{d_0}} \mathord{\left/
 {\vphantom {{{d_0}} d}} \right.
 \kern-\nulldelimiterspace} d}} \right)}^\alpha }} {g_R}$, where $\zeta_0=-30dB$ is the path loss at the reference point $d_0=1\,m$, $d$ denotes the distance between the source and the destination, $\alpha$ shows the path loss exponent which is considered as 3, and ${g_R}$ denotes the small scale fading component. Also, to model the small scale fading, we utilize Rician fading model with Rician factor $K=1$ \cite{3}. Furthermore, we consider noise power as $\sigma^2_n=-90$ dBm.
Moreover, we assume that ${\bf{w}}_p=\sqrt{P_p}\frac{{\bf{h}}_p}{||{\bf{h}}_p||}$, where $P_p$ is the maximum total power of the PBS. Also, we use CVX toolbox of MATLAB to solve the resultant convex problems, and adopt the Monte Carlo method to obtain the results. 

\begin{figure}[!t]
\centering
\begin{minipage}[b]{0.6\textwidth}
\includegraphics[width=\textwidth]{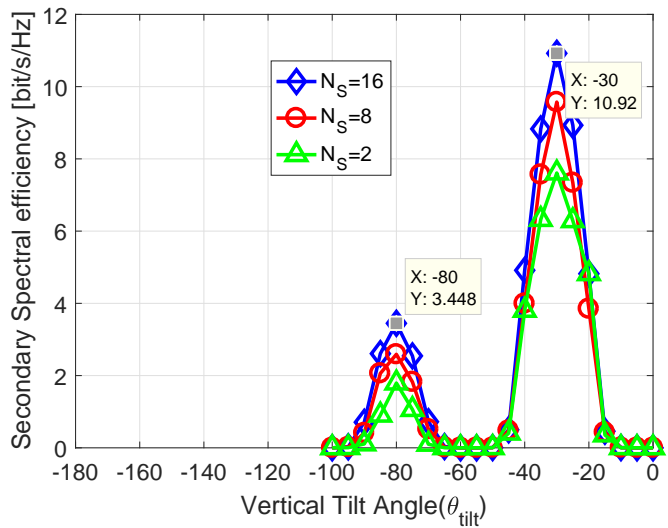}
 \caption{\label{Fig2} Secondary spectral efficiency versus the vertical tilt angle ($\theta_{tilt}$)  in different values of the number of antennas at the SBS ($N_s$) (with $P=10 dBw$ and $N=20$).}
\end{minipage}
\end{figure}
\begin{figure}[!t]
\centering
\begin{minipage}[b]{0.6\textwidth}
   \includegraphics[width=\textwidth]{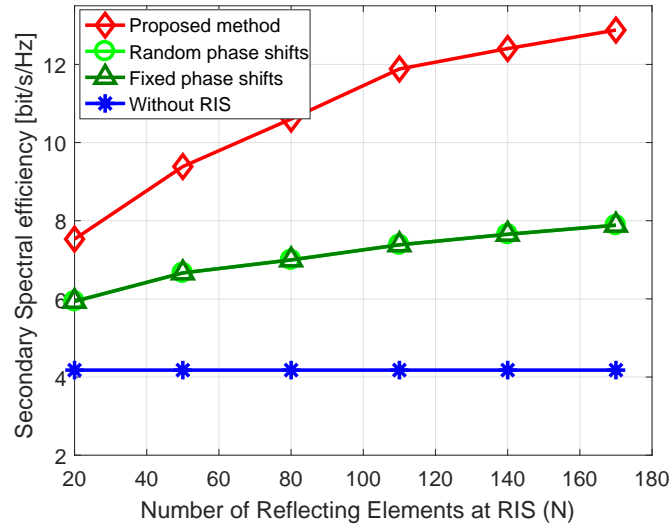}
 \caption{\label{Fig3} Secondary spectral efficiency versus the number of reflecting elements at the RIS ($N$) (with $P=10 dBw$, $N_s=2$ and optimized tilt angle).}
\end{minipage}
\end{figure}
\begin{figure}[!t]
\centering
\begin{minipage}[b]{0.6\textwidth}
   \includegraphics[width=\textwidth]{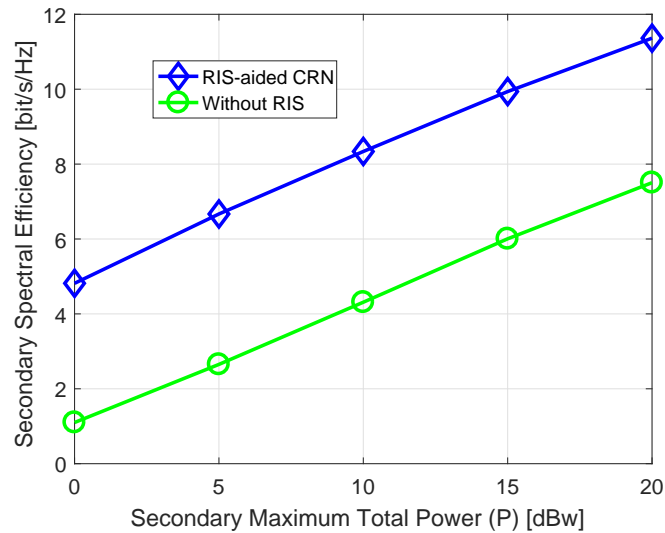}
 \caption{\label{Fig4} Secondary spectral efficiency versus the secondary maximum allowed total power ($P$) (with  $N_s=4$, $N=20$ and optimized tilt angle).}
\end{minipage}
\end{figure}
Fig. \ref{Fig2} shows the secondary spectral efficiency versus the  vertical tilt angle ($\theta_{tilt}$). 
 It can be seen that when the vertical tilt angle ($\theta_{tilt}$) is adjusted to be equal to the elevation angle of the RIS ($\theta_{r}$), the maximum SE at the secondary system is achieved. In other words, the optimum value of $\theta_{tilt}$ is equal to  $\theta_{r}$. Also, another peak in the secondary SE occurs at $\theta_{d}$ shows that when there is no RIS in the network, the optimum value of $\theta_{tilt}$ is equal to  $\theta_{d}$. However, in this case, the achievable SE is lower than the case that RIS is used. Besides, we can see that increasing the number of antennas at the SBS ($N_s$) enhances the secondary spectral efficiency. 

Fig. \ref{Fig3} depicts the secondary spectral efficiency versus the number of reflecting elements of the RIS ($N$) at the optimum value of $\theta_{tilt}$ that obtains in Fig. \ref{Fig2}. It is observed that the SE of the proposed RIS phase optimization method is higher than a scheme with random and fixed phase shifts at the RIS. Also, it is observed that the SE always increases with $N$. Moreover, by increasing $N$, the gap between the ``proposed method'', and other methods becomes larger, which shows that our proposed method is particularly effective for larger $N$. 

Fig. \ref{Fig4} illustrates that the spectral efficiency of the secondary system improves when the secondary system has a large maximum allowed total power ($P$). In addition, we can see that the system equipped with the RIS has better performance than the system with no RIS at all SBS powers. 
\section{Conclusion}\label{Conclusion}
In this letter, we investigated vertical beamforming in an RIS-aided CRN. We formulated a maximization problem to improve the secondary spectral efficiency, and then proposed  an efficient method to solve it. Numerical results showed that the network performance improves when the SBS  orientation is towards the RIS and also when the RIS is equipped with a large number of reflecting elements.


\begin{thebibliography}{30}
\bibitem{1}
 Q. Wu and R. Zhang, “Intelligent reflecting surface enhanced wireless network via joint active and passive beamforming,” IEEE Trans. Wireless Commun., vol. 18, no. 11, pp. 5394-5409, Nov. 2019.
\bibitem{3}
M. Cui, G. Zhang and R. Zhang, "Secure Wireless Communication via Intelligent Reflecting Surface," IEEE Wireless Commun. Lett., vol. 8, no. 5, pp. 1410-1414, Oct. 2019.
\bibitem{4}
J. Ye, S. Guo and M. Alouini, "Joint Reflecting and Precoding Designs for SER Minimization in Reconfigurable Intelligent Surfaces Assisted MIMO Systems," IEEE Trans. Wireless Commun., vol. 19, no. 8, pp. 5561-5574, Aug. 2020.
\bibitem{40}
M. Di Renzo et al., "Smart Radio Environments Empowered by Reconfigurable Intelligent Surfaces: How It Works, State of Research, and The Road Ahead," IEEE J. Sel. Areas Commun., vol. 38, no. 11, pp. 2450-2525, Nov. 2020.
\bibitem{05}
 Q. Wu and R. Zhang, “Towards smart and reconfigurable environment: Intelligent reflecting surface aided wireless network,” IEEE Commun. Mag., vol. 58, no. 1, pp. 106-112, Jan. 2020.
\bibitem{13_new}
F. Hu, B. Chen and K. Zhu, "Full Spectrum Sharing in Cognitive Radio Networks Toward 5G: A Survey," IEEE Access, vol. 6, pp. 15754-15776, Feb. 2018.
\bibitem{06}
L. Zhang, Y. Wang, W. Tao, Z. Jia, T. Song and C. Pan, "Intelligent Reflecting Surface Aided MIMO Cognitive Radio Systems," IEEE Trans. Veh. Technol., vol. 69, no. 10, pp. 11445-11457, Oct. 2020.
\bibitem{60}
J. Yuan, Y. Liang, J. Joung, G. Feng and E. G. Larsson, "Intelligent Reflecting Surface (IRS)-Enhanced Cognitive Radio System," ICC 2020 - 2020 IEEE International Conference on Communications (ICC), Dublin, Ireland, 2020, pp. 1-6.
\bibitem{6}
J. Yuan, Y. -C. Liang, J. Joung, G. Feng and E. G. Larsson, "Intelligent Reflecting Surface-Assisted Cognitive Radio System," IEEE Trans. Commun., vol. 69, no. 1, pp. 675-687, Jan. 2021.
\bibitem{70}
D. Xu, X. Yu, Y. Sun, D. W. K. Ng and R. Schober, "Resource Allocation for IRS-assisted Full-Duplex Cognitive Radio Systems," IEEE Trans. Commun, vol. 68, no. 12, pp. 7376-7394, Dec. 2020.
\bibitem{7}
 X. Guan, Q. Wu, and R. Zhang, “Joint power control and passive beamforming in IRS-assisted spectrum sharing,” IEEE Commun. Lett., vol. 24, no. 7, pp. 1153–1157, Jul. 2020.
\bibitem{2}
X. Tan, Z. Sun, J. M. Jornet, and D. Pados, “Increasing indoor spectrum sharing capacity using smart reflect-array,” in Proc. 2016 IEEE Int. Conf. Commun. (ICC), Kuala Lumpur, Malaysia, May. 2016, pp. 1–6.
\bibitem{8}
H. Xiao, L. Dong, and W. Wang, “Intelligent reflecting surface-assisted secure multi-input single-output cognitive radio transmission,” Sensors (Switzerland), vol. 20, no. 12, pp. 1–23, Jun. 2020.
\bibitem{12_new}
S. M. Razavizadeh, M. Ahn and I. Lee, "Three-Dimensional Beamforming: A new enabling technology for 5G wireless networks," IEEE Signal Process. Mag., vol. 31, no. 6, pp. 94-101, Nov. 2014.
\bibitem{111}
W. Lee, S. Lee, H. Kong, S. Lee and I. Lee, "Downlink Vertical Beamforming Designs for Active Antenna Systems," IEEE Trans. Commu., vol. 62, no. 6, pp. 1897-1907, Jun. 2014.
\bibitem{5}
Q. Nadeem, A. Kammoun, and M. Alouini, “Elevation Beamforming With Full Dimension MIMO Architectures in 5G Systems: A Tutorial,” IEEE Commun. Surveys \& Tutorials, vol. 21, no. 4, pp. 3238-3273, Jul. 2019.
\bibitem{9}
S. M. Razavizadeh and T. Svensson, "3D Beamforming in Reconfigurable Intelligent Surfaces-assisted Wireless Communication Networks," WSA 2020; 24th International ITG Workshop on Smart Antennas, Hamburg, Germany, 2020, pp. 1-5.
\bibitem{14}
P. Cao, J. Thompson, and H. V. Poor, “A sequential constraint relaxation algorithm for rank-one constrained problems,” in Proc. Eur. Signal Process. Conf. (EUSIPCO), pp. 1060-1064, 2017.
\end{thebibliography}
\end{document}